\documentstyle[11pt]{report}
\begin{document}
\newtheorem{thm}{Theorem}
\newtheorem{lemma}[thm]{Lemma}
\newtheorem{propo}[thm]{Proposition}

\centerline {\Large \bf On the Classicality of Broda's SU(2)
Invariants of 4-Manifolds}
\bigskip
\medskip

\centerline {\large Louis Crane\footnote{Supported by National Science
Foundation grant \#DMS-9106476}}
\smallskip

\centerline {\small \it Department of Mathematics}
\centerline {\small \it Kansas State University}
\centerline {\small \it Manhattan, KS 66506-2602}
\bigskip

\centerline {\large Louis H. Kauffman\footnote{Supported by National Science
Foundation grant \#DMS-9205277 and the Program for Mathematics and
Molecular Biology of the University of California at Berkeley, Berkeley, CA}}
\smallskip

\centerline {\small \it Department of Mathematics, Statistics and
	Computer Science}
\centerline {\small \it University of Illinois at Chicago}
\centerline {\small \it Chicago, IL 60680}
\bigskip

\centerline {\large David N. Yetter}
\smallskip

\centerline {\small \it Department of Mathematics}
\centerline {\small \it Kansas State University}
\centerline {\small \it Manhattan, KS 66506-2602}
\bigskip

\noindent{\small {\bf Abstract:} Recent work of Roberts [R] has shown that
the surgical 4-manifold invariant of Broda [B1] and (up to an unspecified
normalization factor) the state-sum 4-manifold invariant of Crane-Yetter [CY]
are equivalent to the
signature of
the 4-manifold.
Subsequently Broda [B2] defined another surgical
invariant of 4-manifolds in which the 1- and 2- handles are treated
differently.
We use a refinement of Roberts' techniques developped in [CKY] to identify
the normalization factor
 to show that the ``improved''
surgical invariant of Broda [B2] also depends only on the signature and
Euler character.}
\bigskip

	As a starting point, let us first observe that the construction of
Crane-Yetter [CY] does not really depend on the use of labels chosen from
the irreps of $U_q(sl_2)$ at the principal $r^th$ root of unity: the simple
objects of any artinian semi-simple tortile category (cf. [S, Y])
in which all objects
are self-dual and the fusion rules are multiplicity free will suffice.
In particular, if we restrict to the integer spin (bosonic)\footnote{This use
of bosonic is a hideous abuse of language--everything in sight
has braid statistics--the ``bosons'' of this paper are the result of
q-deforming honest bosons.} irreps, we obtain
a construction of a different invariant of 4-manifolds.

	In what follows, we use Temperley-Lieb recoupling theory
(cf. [KL,L,R]). In particular,
arcs are labelled with elements of
$\{0,1,...r-2\}$ (twice the spin), $A = e^{2\pi i/4r}$, $q = A^2$,
$\Delta(n) = (-1)^n\frac{q^{n+1} - q^{-n-1}}{q - q^{-1}}$, $\theta(a,b,c)$
denoted the evaluation of the theta-net with edge labelled $a, b,$ and $c$,
and $15-j$ denotes the evaluation of the Temperley-Lieb version of the
Crane-Yetter quantum 15$j$-symbol (with indices suppressed).

	We then adopt the following further notational conventions:

	Arcs labelled $\omega $ denote the linear combination of
arcs labelled $0,1,...,r-2$ in which the coefficient of $i$ is $\Delta(i)$.
Arcs labelled $\tilde{\omega }$ denote the linear combination of arcs
labelled $0,2,...,2\lfloor \frac{r-2}{2} \rfloor$ (even integers) in which
the coefficient of $i$ is $\Delta(i)$.  $N$ denotes the sum of the
squares of the $\Delta(i)$'s, $\tilde{N}$ denotes the sum of the
squares of the $\Delta(i)$'s for $i$ even. Let $\kappa $ be as in
[KL,R], the evaluation of an $\omega $ labelled 1-framed unknot divided by
the positive square root of $N$, and let $\tilde{\kappa }$ be the evaluation
of an $\tilde{\omega }$ labelled 1-framed unknot divided by $\tilde{N}$.

	If $L$ is a framed link, then $\tilde{\omega }(L)$ denotes the
evalutation of the link with all components labelled $\tilde{\omega }(L)$
If $\cal L$ is a set of 4-manifold surgery instructions (cf. Kirby [K]),
 that is a link $L$
with a distinguished 0-framed unlink $\dot{L}$,
then ${\cal B}^!({\cal L})$ denotes
the evaluation of the link $L$ with all components of $\dot{L}$ (one-handle
attachments) colored
$\omega $ and all other components of $L$ (two-handle attachments) colored
$\tilde{\omega }$.

	We then have

\begin{lemma}
	$\tilde{\omega }(L)$ is invariant under handle-sliding.
${\cal B}^!{\cal L}$ is invariant under handle-sliding of 1- and 2-handles
1-handles and of 2-handles over 2-handles.
\end {lemma}

\noindent {\bf proof:}  This follows immediately from handle-sliding over
components labelled $\omega $ and the analysis given
in Remark 17 \S 12.6 of Kauffman/Lins [KL] once it is observed that
pairs of bosons only couple to produce bosons. $\Box$
\smallskip

	And

\begin{lemma}
	{\bf (The bosonic encirclement lemma)}

\begin{figure}[h]
\centering

\setlength{\unitlength}{0.0125in}%
\begin{picture}(285,110)(35,600)
\thicklines
\put(180,675){\line(-1, 0){ 10}}
\put(170,635){\line( 1, 0){ 15}}
\put(200,675){\line(-1, 0){ 10}}
\multiput(170,675)(-0.40000,-0.40000){26}{\makebox(0.4444,0.6667){\sevrm .}}
\put(160,665){\line( 0,-1){ 20}}
\multiput(160,645)(0.40000,-0.40000){26}{\makebox(0.4444,0.6667){\sevrm .}}
\put(170,635){\line( 1, 0){ 15}}
\multiput(200,675)(0.40000,-0.40000){26}{\makebox(0.4444,0.6667){\sevrm .}}
\put(210,665){\line( 0,-1){ 20}}
\multiput(210,645)(-0.40000,-0.40000){26}{\makebox(0.4444,0.6667){\sevrm .}}
\put(200,635){\line(-1, 0){ 15}}
\put(115,675){\line(-1, 0){ 10}}
\put(105,635){\line( 1, 0){ 15}}
\put(135,675){\line(-1, 0){ 10}}
\multiput(105,675)(-0.40000,-0.40000){26}{\makebox(0.4444,0.6667){\sevrm .}}
\put( 95,665){\line( 0,-1){ 20}}
\multiput( 95,645)(0.40000,-0.40000){26}{\makebox(0.4444,0.6667){\sevrm .}}
\put(105,635){\line( 1, 0){ 15}}
\multiput(135,675)(0.40000,-0.40000){26}{\makebox(0.4444,0.6667){\sevrm .}}
\put(145,665){\line( 0,-1){ 20}}
\multiput(145,645)(-0.40000,-0.40000){26}{\makebox(0.4444,0.6667){\sevrm .}}
\put(135,635){\line(-1, 0){ 15}}
\put( 70,670){\line( 0, 1){  5}}
\put( 35,675){\line( 1,-1){ 20}}
\put( 70,675){\line(-1, 0){ 30}}
\put( 70,640){\line( 0,-1){  5}}
\put( 35,635){\line( 1, 1){ 20}}
\put( 70,635){\line(-1, 0){ 30}}
\put( 35,675){\line( 1, 0){  5}}
\put(115,675){\line( 1, 0){ 10}}
\put(185,710){\line( 0,-1){ 70}}
\put(185,630){\line( 0,-1){ 30}}
\put( 35,635){\line( 1, 0){  5}}
\put( 40,615){\makebox(0,0)[lb]{\raisebox{0pt}[0pt][0pt]{\twlrm j even}}}
\put(115,620){\makebox(0,0)[lb]{\raisebox{0pt}[0pt][0pt]{\twlrm j}}}
\put(215,650){\makebox(0,0)[lb]{\raisebox{0pt}[0pt][0pt]{\twlrm j}}}
\put(190,690){\makebox(0,0)[lb]{\raisebox{0pt}[0pt][0pt]{\twlrm n}}}
\put(265,650){\makebox(0,0)[lb]{\raisebox{0pt}[0pt][0pt]{\twlrm =}}}
\put(320,645){\makebox(0,0)[lb]{\raisebox{0pt}[0pt][0pt]{\twlrm 0}}}
\end{picture}

\end{figure}

\noindent whenever $n$ is even and non-zero.
\end{lemma}

\noindent {\bf proof:} This follows from the same proof as the encirclement
lemma of Lickorish [L] (cf. also Kauffman/Lins [KL]) with the ``auxiliary
loop'' labelled $2$ instead of $1$. $\Box$

	Let
{\small
\[ CY_B(W) = \tilde{N}^{n_0-n_1}\sum_{\parbox{.65in}{\tiny
	\begin{center}even
labellings $\lambda$
of faces and tetrahedra \end{center} }}
	\prod_{\parbox{.25in}{\tiny
\begin{center} faces $\sigma $ \end{center} }}
\Delta(\lambda(\sigma ))\prod_{\parbox{.5in}{\tiny
\begin{center} tetrahedra $\tau $ \end{center} }}
\frac{\Delta(\lambda(\sigma ))}{ \theta(\lambda(\tau ), \lambda(\tau _0),
\lambda(\tau _2)) \theta(\lambda(\tau ), \lambda(\tau _1),
\lambda(\tau _3))}\prod_{\parbox{.6in}{\tiny
\begin{center} 4-simplexes \end{center} }}\! 15-j \]
}

be the bosonic Crane-Yetter invariant.

	Let $|L|$ (resp. $\nu (L)$, $\sigma (L)$) denote the number of
components of a link $L$ (resp. the nullity of the linking matrix of $L$,
the signature of the linking matrix of $L$).

	We can then define a purely bosonic version of Broda's original
invariant by

\[ Br_B(W) = \frac{\tilde{\omega }(L)}{\tilde{N}^\frac{|L| + \nu(L)}{2}} \]

\noindent where $L$ is the underlying link of a surgery presentation of
$W$; while a bosonic version of the Reshetikhin/Turaev [RT] 3-manifold
invariant is given by

\[ I_B(M) = \tilde{\kappa }^{-\sigma(L)}\tilde{N}^\frac{|L| + 1}{2}
	\tilde{\omega }(L) \]

\noindent where $L$ is a framed link giving surgery instructions for $M$.

	Applying the two lemmas above in an analysis otherwise
identical to that of given by Roberts [R] of the
original Broda invariant [B1] shows that

\begin{propo}

\[ Br_B(W) = \tilde{\kappa }^{\sigma(W)} \]

\end{propo}

	Similarly it follows from the bosonic encirclement lemma that

\[ CY_B(W) = \tilde{N}^{n_0-n_1-n_3}\tilde{\omega }(L) \]

\noindent where $n_d$ is the number of $d$-simplexes in a triangulation, and
$L$ is the link derived from a triangulation by putting a 0-framed unknot
in each tetrahedron, and a loop around each 2-simplex (running mostly
through 4-simplexes but linking each tetrahedron's unknot) after the manner
of Roberts [R].

	It then follows as in [CKY] that

\begin{propo}

\[ CY_B(W) = \tilde{\kappa }^{\sigma (W)}\tilde{N}^\frac{\chi(W)}{2}
	\hspace*{.3in} \parbox{.3in}{(*)} \]

\end{propo}

	Now, Broda's new invariant is defined by

\[ {\cal B}(W) =
	\frac{{\cal B}^!(L)}{\tilde{N}^{\nu (L)}N^\frac{|L|-\nu (L)}{2}} \]

	For convenience we first analyse a slightly different
normalization (for which the proof of invariance is effectively
identical to that for {\cal B}(W): let

\[ {\bf B}(W) =
	\frac{{\cal B}^!(L)}{\tilde{N}^{|L-\dot{L}| - |\dot{L}|}
	N^{|\dot{L}|}}  \]

	Now, it follows from the original encirclement lemma of Lickorish [L]
that

\[ CY_B(W) = \tilde{N}^{n_0-n_1}N^{-n_3}{\cal B}^!({\cal L})
	\hspace*{.3in} \parbox{.3in}{(**)}\]

\noindent where $\cal L$ is the surgery instructions given by assiociating
the link $L$ to the triangulation as above, and letting $\dot{L}$ be the
unlink of loops in the tetrahedra.

	Observe that {\bf B} is multiplicative under connected sum, and
that ${\bf B}(S^1 \times S^3) = \tilde{N}$ (an easy calculation).
As shown in Roberts [R],
$\cal L$ is a surgery presentation for
$W\#(\stackrel{n_4-1}{\#} S^1 \times S^3)$.

	From this and the fact that for $\cal L$, $|L - \dot{L}| = n_2$ and
$|\dot{L}| = n_3$, we see that

\begin{eqnarray*}
 \frac{{\cal B}^!({\cal L})}{\tilde{N}^{n_2-n_3}N^{n_3}} &  = &
	{\bf B}(W\#(\stackrel{n_4-1}{\#} S^1 \times S^3)) \\
		& = & {\bf B}(W)\tilde{N}^{n_4-1}. \\
\end{eqnarray*}

	Thus

\[ {\cal B}^!(L) = {\bf B}(W) \tilde{N}^{n_2-n_3+n_4-1} N^{n_3}.
	 \hspace*{.3in} \parbox{.4in}{(***)}\]

	It then follows from (*), (**) and (***) that

\[ {\bf B}(W) =
	\tilde{\kappa }^{\sigma(W)}\tilde{N}^{\frac{\chi (W)}{2} - 1} \]

	To return to Broda's [B2] original normalization, note that

\[ {\cal B}(W) =
	{\bf B}(W)(\tilde{N} N^{-\frac{1}{2}})^{|L-\dot{L}|-|\dot{L}|-\nu (L)}
	\]

	From which we obtain

\begin{thm} If $W$ is a connected closed oriented smooth 4-manifold, then

\[	{\cal B}(W) =
	\tilde{\kappa }^{\sigma (W)}
	\left( \frac{\tilde{N}}{N} \right) ^{\frac{\chi (W)}{2}-1}  \]
\end{thm}

\noindent {\bf proof:} It suffices to shown that if $W$ is given by the
surgery instruction $\cal L$, then

\[ |L-\dot{L}|-|\dot{L}|-\nu (L) = \chi (W) - 2. \]

But this follows immediately from the observation that $\nu (L)$ is
the number of 3-handles attached in completing the construction of $W$. $\Box$

\centerline {\bf References}
\bigskip

\noindent [B1] Broda, B., {\em Surgical invariants of 4-manifolds}, preprint
(1993).
\smallskip

\noindent [B2] Broda, B., {\em A Surgical invariant of 4-manifolds},
Proceedings of the Conference on Quantum Topology (D.N. Yetter, ed.),
World Scientific, to appear.
\smallskip

\noindent [CKY] Crane, L., Kauffman, L. H. and Yetter, D. N., {\em Evaluating
the Crane-Yetter invariant}, e-preprint hep-th/9309063, and to appear
{\em Quantum Topology} (R. Baadhio and L.H.
Kauffman eds.), World Scientific.
\smallskip

\noindent [CY] Crane, L. and Yetter, D. N., {\em A categorical construction
of 4D topological quantum field theories}, e-preprint
hep-th/9301062, and to appear in {\em Quantum Topology} (R. Baadhio and L.H.
Kauffman eds.), World Scientific.
\smallskip

\noindent [KL] Kauffman, L. H. and Lins, S. L. {\em Temperley-Lieb Recoupling
Theory and Invariants of 3-Manifolds}, Princeton University Press, to appear.
\smallskip

\noindent [K] Kirby, R., {\em The topology of 4-manifolds}, SLNM vol. 1374,
Springer-Verlag (1989).
\smallskip

\noindent [L] Lickorish, W. B. R. {\em The Temperley-Lieb algebra and
3-Manifold invariants}, preprint (1992).
\smallskip

\noindent [RT] Reshetikhin, N. and Turaev, V. G., {\em Invariants of
3-manifolds via link polynomials and quantum groups}, Invent. Math.
{\bf 103} (1991), 547-597.
\smallskip

\noindent [R] Roberts, J, {\em Skein theory and Turaev-Viro invariants},
preprint (1993).
\smallskip

\noindent [S] Shum, M.-C., {\em Tortile tensor categories}, Doctoral
Dissertation, Macquarie University, 1989.
\smallskip

\noindent [Y] Yetter, D.N., {\em State-sum invariants of 3-manifolds
associated to artinian semisimple tortile categories}, Topology and its
App., to appear.
\end{document}